\documentstyle[twocolumn,prb,aps,epsf]{revtex}

\begin{document}

\draft
\title{
Large Skyrmions in an Al$_{0.13}$Ga$_{0.87}$As Quantum Well
}
\author{S. P. Shukla, M. Shayegan, S. R. Parihar, S. A. Lyon
}
\address{
Department of Electrical Engineering, Princeton University,
Princeton,
New Jersey 08544
}
\author{N. R. Cooper
}
\address{School of Physics and Astronomy, University
of Birmingham, Edgbaston, Birmingham B15 2TT, United Kingdom
}
\author{A. A. Kiselev
}
\address{Department of Electrical and Computer Engineering,
North Carolina State University,\\ Raleigh, North Carolina 27695-7911
}
\date{\today}
\maketitle
\begin{abstract}

We report tilted-field magnetotransport measurements of
two-dimensional electron systems in a 200 \AA-wide
Al$_{0.13}$Ga$_{0.87}$As quantum well.  We extract the energy gap for
the quantum Hall state at Landau level filling $\nu =1$ as a
function of the tilt angle. The relatively small effective Land\'{e}
$g$-factor ($g \simeq 0.043$) of the structure leads to skyrmionic
excitations composed of the largest number of spins yet reported ($s
\simeq 50$). Although consistent with the skyrmion size observed,
Hartree-Fock calculations, even after corrections, significantly
overestimate the energy gaps over the entire range of our data.  

\end{abstract}

\narrowtext


In two-dimensional electron systems (2DESs), the quantum Hall effect
(QHE) at Landau level filling factor $\nu=1$ has attracted much
theoretical \cite{Sondhi93,Fertig94,Wu95,Brey95,Fertig97,Cooper97} and
experimental
\cite{Barrett95,Schmeller,Aifer96,Bayot96,Maude,Leadley98,%
Nicholas98,Bayot99} attention. At this filling, the Coulomb (exchange)
energy is so influential that the QHE excitation gap is
more than 
an order of magnitude larger than the single-particle Zeeman energy,
the gap expected if the
Coulomb energy were ``turned off.''
In fact, given a small enough Zeeman
energy, the interplay between these two energies leads to a
lowest-lying charged excitation, called a {\em skyrmion}, composed of
electrons arranged in a canted, nearly-parallel spin-texture.
Properties of this excitation, such as its energy gap and
physical extent, are determined by the ratio $\tilde{g} = |g|\mu_B
B_{\rm tot}/(e^2/\epsilon \ell_B)$ of the single-particle Zeeman
energy,
which limits the number of spin-flips in an excitation, to the Coulomb
energy which favors local ferromagnetic ordering. ($\epsilon$ is
the dielectric constant, $\ell_B = \sqrt{\hbar c /e B_\perp}$ is the
magnetic length, $g$ is the effective Land\'{e} $g$-factor, $\mu_B$
the Bohr magneton, and $B_{\rm tot}$ and $B_\perp$ are the total
applied magnetic field and the component perpendicular
to the layer plane respectively.) 

                    
The limit of $\tilde{g} \rightarrow 0$ is of particular
interest where, in an ideal system, the excitation gap is predicted to
exist even in the absence of Zeeman energy and the 
skyrmion size diverges --- $s \rightarrow
\infty$.  (We choose $s$ to denote the total spin of a
thermally-activated skyrmion-antiskyrmion pair; the effective spin of
a single skyrmion or antiskyrmion would therefore be $s$/2 if particle
symmetry holds.) Experiments employing diverse techniques such as
optically-pumped nuclear magnetic resonance,\cite{Barrett95}
magnetotransport,\cite{Schmeller,Bayot99} and  magneto-optical
absorption
spectroscopy\cite{Aifer96} in GaAs 2DESs, where $|g| \simeq 0.44$,
have
yielded $s \sim$ 7 to
9.
By using hydrostatic pressure to tune $g$, Refs.
\onlinecite{Maude,Leadley98,Nicholas98} can access the regime
$\tilde{g}
\rightarrow 0$ where
they extract a larger number of spin-flips ($s=36$) from the
temperature dependence of their
magnetotransport data.
Unfortunately,
this technique requires a separate cooldown for each $g$ which leads
to a different disorder potential every time. Furthermore, 
since applying hydrostatic pressure lowers the density of the 2DES, to
compensate, the sample must be illuminated. Because of these
complications, a
controllable, systematic study using pressure is non-trivial.


In this paper, we report the observation of the largest skyrmions yet
reported ($s \simeq 50$) by using an alternate approach to access the
low $\tilde{g}$ regime.  In bulk Al$_x$Ga$_{1-x}$As, $g$ increases
monotonically from $g = -0.44$ at $x=0$ (GaAs) to $g \approx  +0.5$ at
$x = 0.35$ vanishing at $x \simeq 0.13$.\cite{Weisbuch77}  We
fabricated a wafer with a {200 \AA}\ Al$_{0.13}$Ga$_{0.87}$As quantum
well bounded first by a thin 12.6 \AA\ AlAs layer followed by thick
Al$_{0.35}$Ga$_{0.65}$As barriers on each side.  Grown by molecular
beam epitaxy, this symmetric structure is modulation doped with Si.
While many experimental techniques measure
$g$,\cite{Weisbuch77,Dobers88,Snelling91,Heberle94,Sapega94,%
Kalevich92}
determining an extremely low $g$ is difficult and subject to great
relative uncertainty.  From calculations utilizing the Kane
model,\cite{Ivchenko92,Kiselev98} our best estimate of the $g$-factor
for this wafer is $g = 0.043 \pm 0.010$, an order of magnitude lower
than that in bulk GaAs.\cite{gpressure} The slightly positive value of
$g$ is attributed to the spill-over of the electron wavefunction into
the barrier region where $g$ is positive, as well as the
non-parabolicity of the energy bands.\cite{Ivchenko92,Hannak}  In the
right inset to Fig. 1, we show a calculation of the longitudinal
$g$-factor for a symmetrically-distributed electron system as a
function of Al concentration $x$ in a 200 \AA\ Al$_x$Ga$_{1-x}$As
quantum well with barriers as specified above.  We have taken the
$g$-factor to be isotropic; we will address this assumption later. 

We measure two samples (identified as A and B) from different parts of
the wafer with mobility $\mu
\approx 5 \times 10^4\ {\rm cm}^2/{\rm Vs}$ in a Van der
Pauw geometry.  Samples A and B have total areal densities $n = 1.37
\times 10^{11}$ and $1.28 \times 10^{11}$ cm$^{-2}$, respectively.  We
collect the low-temperature magnetotransport data in a dilution fridge
and a $^3$He system. In our experiment, we extract the excitation 
energy gap ($\Delta_1$) for the QHE at filling $\nu = 1$ from the
temperature-dependence of its longitudinal resistance ($R_{xx}$)
minimum.  We
gather
$\Delta_1$ for several
$\tilde{g}$ by tilting the sample and thus applying a magnetic field
$B_{\rm tot}$ at an angle $\theta$ with respect to the normal of the
sample plane.  This technique allows the Zeeman energy ($\propto
B_{\rm tot}$) to be tuned {\em in situ} while the other parameters in
the
system are nearly unaffected.\cite{Schmeller}  Unlike the experiments
in
Refs.\onlinecite{Maude,Leadley98,Nicholas98}, the areal density $n$,
the
disorder, and the Coulomb energy $e^2/{\epsilon \ell_B}$ remain
constant for different values of $\tilde{g}$ leading to a relatively
straightforward analysis of our data.  By finding the energy gap
for several angles $\theta$ (or equivalently $\tilde{g}$), we can
determine the number of spin-flips involved in an excitation since any
change in the gap is almost entirely attributable to the change in the
Zeeman energy contribution.  As in
Ref.\onlinecite{Schmeller} and
Refs.\onlinecite{Maude,Leadley98,Nicholas98}, we use the
formula
$s = \partial\tilde{\Delta}_1/\partial\tilde{g}$, where
$\tilde{\Delta}_1 = \Delta_1 / (e^2/\epsilon \ell_B)$ is $\Delta_1$
normalized by the Coulomb energy, to extract the
number of spin-flips in an excitation.


The traces in Fig.~1 attest to the high quality of the sample as well
as the small value of the $g$-factor.  The longitudinal
magnetoresistance of sample A is plotted for temperatures $T =$~20~mK
and~735 mK.  The 20~mK trace exhibits minima for QHE states with even
integer fillings as high as $\nu = 42$ at $B_\perp = 0.13$ T marking
the lower bound for the Shubnikov-de Haas (SdH) oscillation regime.
(See the left inset to Fig.~1.)   We believe that impurity scattering,
rather than alloy scattering, is the dominant mechanism limiting the
mobility in our AlGaAs quantum wells.  A Born approximation
treatment\cite{chatto85,saxena81} estimates the alloy scattering
mobility limit to be $\approx 4 \times 10^5\ {\rm cm}^2/{\rm Vs}$, an
order of magnitude larger than our measured mobility.  Impurity
scattering, however, may explain the low measured mobility.  It is
known that Al, a relatively reactive element, incorporates
impurities in AlGaAs layers during growth.\cite{Coleridge}  Such a
mechanism is consistent with the relatively low mobility (even after
considering the occupation of multiple ellipsoids) observed in AlAs
quantum wells\cite{sterge99} where there is no alloy scattering.

Although minima corresponding to QHE states exist for many even
integer fillings, there are none for odd integers other than $\nu= 1$
and 3. At higher odd fillings, the influence of the Coulomb exchange
energy is progressively diminished since the states occur at lower
$B_\perp$ and the fraction of electrons affected is
$1/\nu$;\cite{Ando74} instead, the excitation
gap at higher
odd $\nu$ is determined primarily by the competition of the
single-particle Zeeman energy and {disorder-broadening} of Landau
levels.  In our system the Coulomb exchange energy appears to be
significant enough to overcome {disorder-broadening} only for the odd
integer fillings $\nu =$ 1 and 3.  In fact, calculations which
consider the finite width of the 2DES\cite{Cooper97} predict
skyrmionic excitations for our sample at both these fillings.
Unfortunately, the measured excitation gap at $\nu = 3$ is only
$\simeq3$ K which is of the order of the Landau level broadening
\cite{Gamma} making further analysis inconclusive. 


We now discuss our data for $\nu = 1$. Shown in Fig.~2 are the
Arrhenius plots of $R_{xx}$ minima for the $\nu = 1$ QHE in sample A
for three angles in the temperature range 1 $< T <$ 4~K.  We extract
the activation energy $\Delta_1$ from the slope of a best-fit line
(dashed line) to the data using the relation $R_{xx} \sim {\rm
exp}(-\Delta_1 / 2 T)$.  For $\theta =$ 0, 49.7, {\rm and}
67.6$^\circ$, we have $\Delta_1 =$ 15, 20, and 25~K respectively.
Our measured $\Delta_1$ is a monotonically increasing function of
$\theta$ lying in the range {13 $< \Delta_1 <$ 25~K} as shown in the
inset to Fig. 2. Since the data for both samples A and B are
qualitatively very similar, we focus on sample A below.


In Fig. 3, we plot $\tilde{\Delta}_1 = \Delta_1 / (e^2/\epsilon
\ell_B)$ vs $\tilde{g}$ and discuss our results in light of other
experiments and theoretical calculations.  The data from sample A,
which occupies the extreme lower left portion of the figure, is
expanded in the inset to Fig. 3.  An asymptote (dashed line) fit to
the lower range of the data reveals $s = 50.2 \pm 1.0$. Similar
analysis on sample B yields $s = 49.2 \pm 2.1$. 
Compared to our experiment, Ref.\onlinecite{Schmeller} explores higher
$\tilde{g}$ in GaAs samples represented by various closed symbols in
Fig. 3.  
In the case of the pressure-tuned data (not shown) from
Refs.\onlinecite{Maude,Leadley98,Nicholas98}, $\tilde{g}$ is in the
same
vicinity as our data, although there is more scatter in the reported
$\tilde{\Delta}_1$, presumably because of variations in disorder as
discussed before.  A noteworthy commonality in these
experiments is the range of the measured skyrmionic excitation
gap ($\tilde{\Delta}_1 < 0.33)$.

We now contrast our experimentally obtained $\tilde{\Delta}_1$ with
calculations. The top solid curve in Fig. 3 represents the results of
Hartree-Fock calculations \cite{Fertig97,Cooper97} for the skyrmion
excitation energy gap in an ideal, infinitely-thin 2DES, i.e. one with
an electron probability density width $w = 0$. The calculated skyrmion
gap declines steeply as $\tilde{g} \rightarrow 0$, reflecting the
decreasing cost in Coulomb energy for an excitation with an increasing
degree of nearly-parallel spins.  On the other hand, the
(exchange-enhanced) single spin-flip excitation gap (dashed line)
expected in the absence of skyrmions has a constant slope
corresponding to $s = 1$. At $\tilde{g}=0$, the skyrmion gap is $
{1\over2}\sqrt{\pi\over 2} e^2/{\epsilon \ell_B}$, half the single
spin-flip excitation gap. The skyrmion remains the favored excitation
for $\tilde{g} < \tilde{g}_c = 0.054$ (marked by a vertical arrow).

Note the striking discrepancy between the calculations and the
experiments. The calculated skyrmion gap for the ideal case is a
factor of 4.3 to 6.7 larger than the experimental data! The ideal
case, however, ignores important effects such as finite thickness
correction (FTC), Landau level mixing (LLM), and disorder-broadening
of Landau levels
---
all of which reduce the energy gap. Unfortunately, since no
calculation currently treats these three corrections simutaneously for
skyrmions, we must consider them in cumulative succession. Figure 3
includes Hartree-Fock calculations with FTC \cite{Cooper97} for layer
thickness $w = 0.43 \ \ell_B$, appropriate for our sample. (We
determine $w$ by fitting a gaussian function to the electron
probability density from a self-consistent local density approximation
calculation.) By softening short-range interactions, the FTC reduces
the predicted gap by $\approx30\%$. To assess the effect of LLM, we
first focus on the exchange-enhanced single spin-flip excitation gap
corrected for FTC and LLM. Existing calculations are not in
quantitative agreement; based on the trends in
Ref.\onlinecite{Kralik95} and
Ref.\onlinecite{Bonesteel99}, we estimate the corrected single
spin-flip gap
to be about 0.58 and 0.69 ${e^2 /\epsilon \ell_B}$ respectively for
our sample (plus the Zeeman energy, $|g|\mu_B
B_{\rm tot}$).  And, if the role of disorder is limited
to the
disorder-broadening of Landau-levels, we expect the predicted gap to
diminish by only $\Gamma \simeq$ 0.06 ${e^2 /
\epsilon\ell_B}$.\cite{Gamma}  We then deduce the gap in the large
skyrmion limit (small $\tilde{g}$) by
shifting the $w = 0.43 \ \ell_B$ curve for the skyrmion excitation gap
by a constant to match (at $\tilde{g} > \tilde{g}_c$) the corrected
single spin-flip gaps estimated.  The shifted curve still
overestimates our experimental gaps for $\nu = 1$ by a factor of 1.3
to 1.5 and 1.8 to 2.3 for Ref.\onlinecite{Kralik95} and
Ref.\onlinecite{Bonesteel99} respectively.  

Although the {\em absolute} values of the calculation for the $\nu =
1$ QHE cannot be reconciled with the experimental data, the size of
skyrmions (from the slope of the curve) predicted by the calculation
agrees with our data.  This agreement is evidenced by the $w = 0.43 \
\ell_B$ curve shifted down by $0.46 \ {e^2 / \epsilon\ell_B}$ (shown
by the dotted line in the inset to Fig.~3) which fits the entire lower
range of our data remarkably well.  In fact, we can use the
calculation to check the validity of our value for the $g$-factor.  We
find that this agreement is valid only in a very narrow range of
assumed values for $g$ which includes our estimate of
$g \simeq
0.043$.\cite{Cooper_fit}  This congruity may be viewed, perhaps, as an
independent confirmation of $g$-factor in our sample.


Thus far, we have interpreted our data assuming an isotropic
$g$-factor for our sample.  In general, however, the $g$-factor can be
anisotropic in confined systems, with $g_\ell$ and $g_t$ denoting the
longitudinal and transverse components of the $g$-factor with respect
to the growth axis (for a review, see Ref.\onlinecite{Kiselev98} and
references therein).  The electron $g$-factor anisotropy is governed
by the low-symmetry electron quantum confinement and
changes strongly with the quantum well width; it can be qualitatively
estimated from the energy splitting between the
light- and heavy-hole bands. Based on a time-resolved
photoluminescence experiment measuring
electron
spin quantum beats,\cite{Le_Jeune97} Le Jeune {\em et al.} conclude
that the $g$-factor is indeed anisotropic for narrow GaAs quantum
wells bounded by Al$_{0.30}$Ga$_{0.70}$As barriers, as found in
Refs.\onlinecite{Kalevich92,Kalevich95}; however, in quantum wells
120\ \AA\
and wider, the anisotropy vanishes.  We note here that the $g$-factor
anisotropy may be
reduced as the electron kinetic energy in the 2DES, the thermal
energy, or localization energy (because of imperfections or magnetic
field)
become comparable in value to the quantum confinement energy.

A simple Kane-model calculation for electrons at the bottom of the
first subband in our system yields a transverse component of the
$g$-factor $g_t = 0.085$. And, if we reinterpret our data so that
$g(\theta) = \sqrt{g_\ell^2 {\rm cos}^2(\theta) + g_t^2 {\rm
sin}^2(\theta)}$, then $s$ from the asymptote to the lower range of
the data is reduced --- $s = 19$. However, experimental findings
mentioned above support an isotropic $g$-factor for the parameters in
our sample.  Our system with a quantum well of width 200\ \AA\ and a
finite 2DES concentration should exhibit even less of a tendency
toward
$g$-factor anisotropy. And, as already noted, the size of
skyrmions
from the calculations and our data are no longer consistent if the
anisotropic $g$-factor is assumed.  Therefore, we believe that the
isotropic value of 0.043 is the best estimate for the $g$-factor in
our samples. 


In summary, we have focused on the thermal excitation energy for $\nu
= 1$ gathered for several tilt-angles from magnetotransport
measurements of two-dimensional electron systems in a 200 \AA-wide
Al$_{0.13}$Ga$_{0.87}$As quantum well. In this structure with a small
$g$-factor ($g \simeq 0.043$), we observe skyrmions of the largest
size yet
reported ($s \simeq 50$). The magnitude of the energy gaps measured
are consistent with those from other experiments. And, while matching
the experimentally determined size of skyrmions, Hartree-Fock
calculations, even after treatment for corrections, significantly
overestimate the energy gaps in our data. Understanding this disparity
requires further studies in the large skyrmion regime.

We acknowledge discussions with D. C. Tsui, S. L. Sondhi, K. Lejnell,
N. E. Bonesteel, and H. A. Fertig and the use of NHMFL in Tallahassee,
Florida.  This work was supported by the National Science Foundation
grants DMR-9809483 and DMR-9971021.  A.A.K. was supported, in part, by
the Office of Naval Research.

%
%

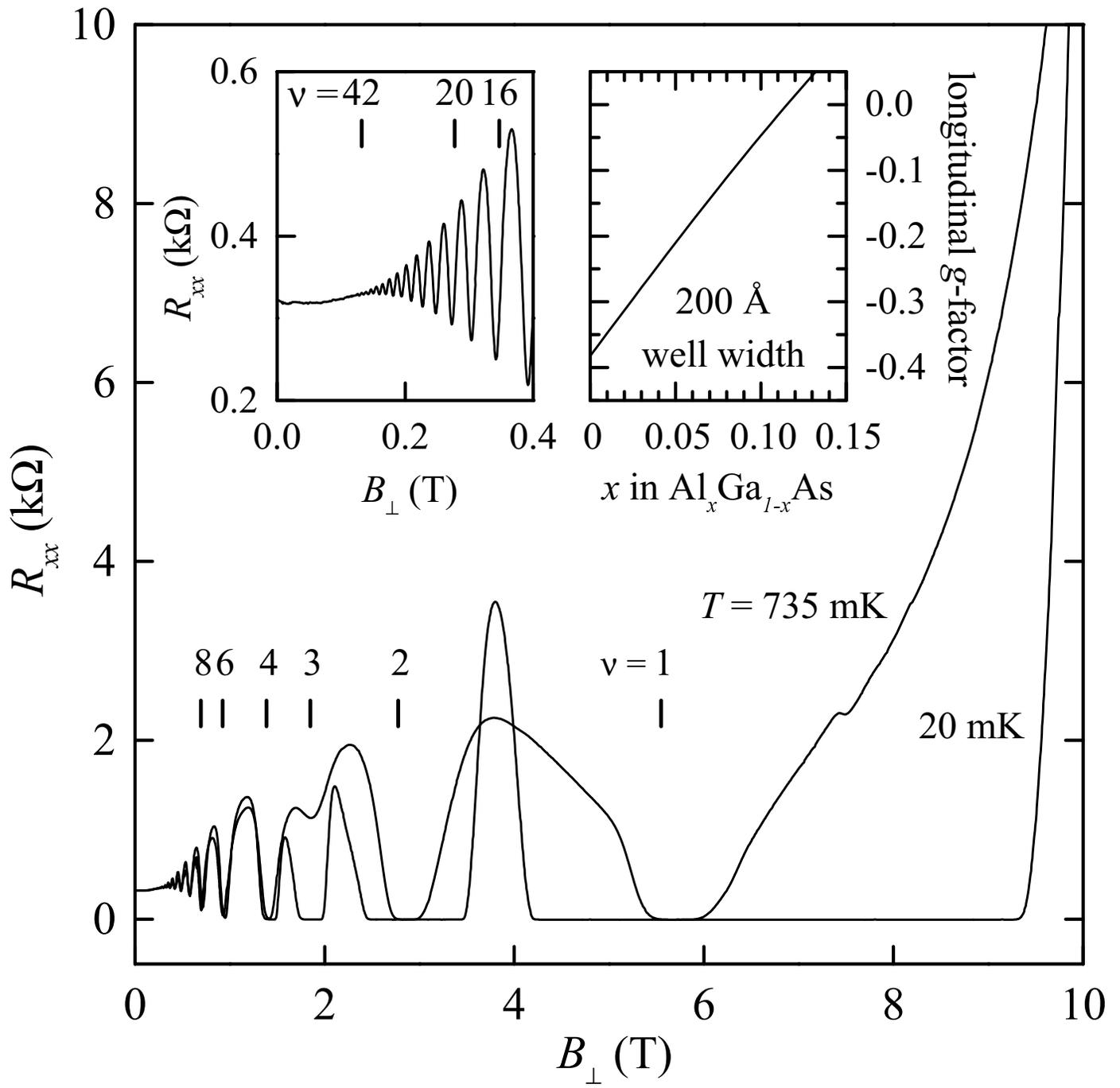
\begin{figure}
\caption{Magnetoresistance traces for sample A at $T = 20$ and 735 mK
which show QHE states
for only $\nu = 1$ and 3 among odd $\nu$. The left inset shows the
onset of
Shubnikov-de Haas oscillation at $\nu = 42$ ($B_\perp = 0.13$ T).  The
right
inset shows a calculation of the longitudinal $g$-factor
$g_\ell$ for a 200 \AA\ Al$_x$Ga$_{1-x}$As well bounded by
Al$_{0.35}$Ga$_{0.65}$As; $g_\ell = 0.043 \pm 0.010$ for $x = 0.13$.
}
\label{fig1}
\end{figure}

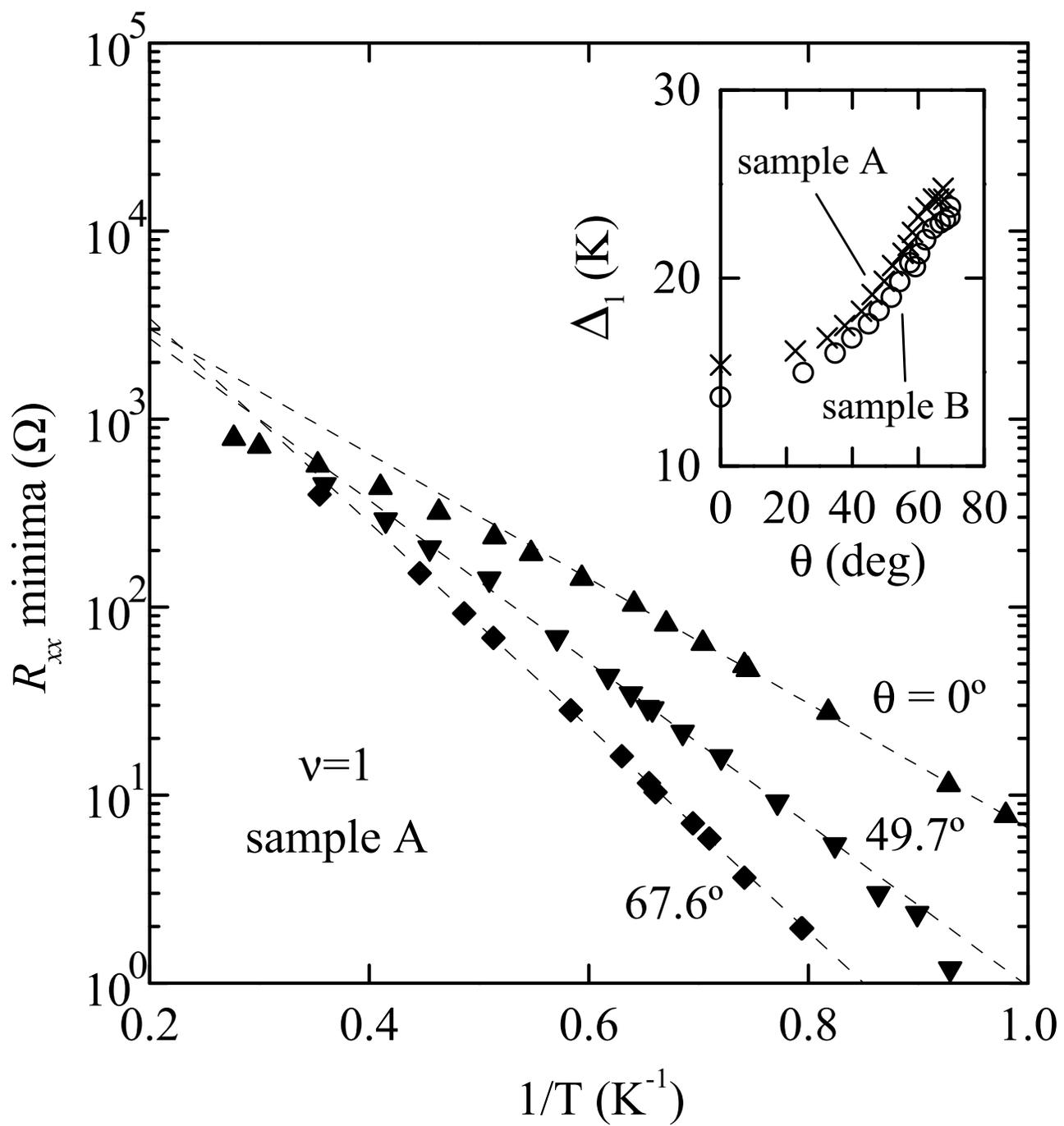
\begin{figure}
\caption{Arrhenius plot of $R_{xx}$ minima for
the $\nu = 1$ QHE state at several angles $\theta$ for sample A.
The activation energy $\Delta_1$ is plotted in the inset vs
$\theta$ for both samples A and B.  
}
\label{fig2}
\end{figure}

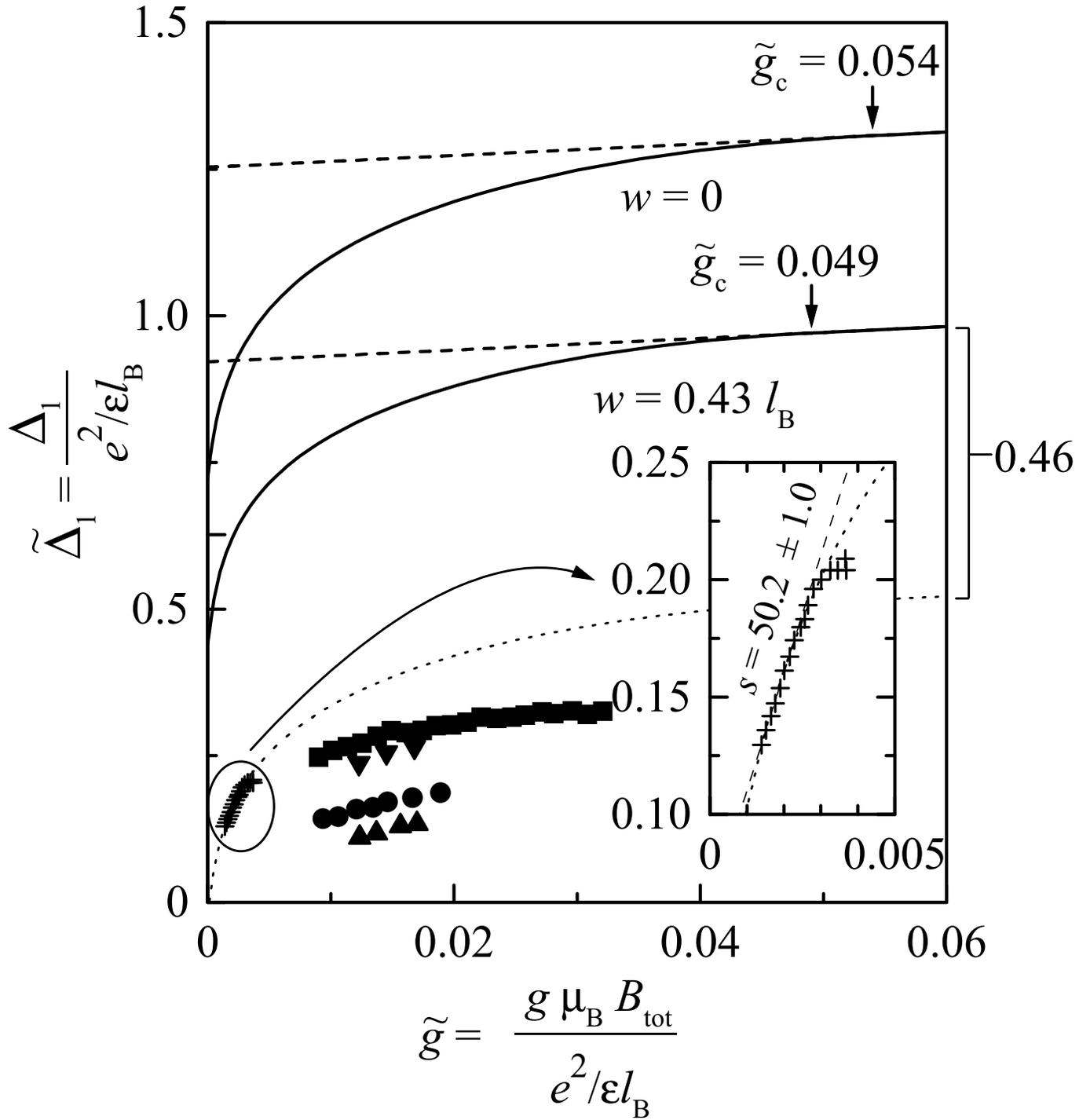
\begin{figure}
\caption{Normalized activation energy $\tilde{\Delta}_1 =
\Delta_1 /(e^2/\epsilon \ell_B)$ vs normalized Zeeman energy
$\tilde{g} = g \mu_B B_{\rm tot} / (e^2/ \epsilon \ell_B)$ from
experiments and calculations.  The experimental data are from sample A
($+$) and Ref.~8 (closed symbols).  In the inset, the asymptote
(dashed line) fit to the lower range of sample A data reveals $s =
50.2 \pm 1.0$. The results of Hartree-Fock calculations for a 2DES
with zero layer-thickness ($w = 0$) and for $w = 0.43\ \ell_B$ are
also shown in the main figure.  In the inset, the  $w = 0.43 \ \ell_B$
skyrmion excitation gap (dotted line) shifted down by $ 0.46\thinspace
e^2 / \epsilon \ell_B$ matches the lower range of sample A data.  (See
text
for details.)
}
\label{fig3}
\end{figure}


\begin{thebibliography}{10}
\bibitem{gpressure}
We note that the pressure-tuned efforts (Refs.~11--13) too must use a
  calculation to convert the hydrostatic pressure to $g$.
\bibitem{Cooper_fit}
We note that the agreement between the data and the calculation is
visibly
  worsened if an isotropic $g$ outside $0.04 < g < 0.05$ is assumed.
\end{thebibliography}
\end{document}